\documentclass[useAMS,usenatbib]{mn2e}

\usepackage{graphicx}
\usepackage{setspace}
\usepackage{natbib}
\usepackage{color}
\usepackage{amsmath,amssymb}
\usepackage{times}
\usepackage{aas_macros}
\usepackage{multirow}

\title[The cold veil of the Milky Way stellar halo]{The cold veil of the Milky
  Way stellar halo\thanks{Based on observations made with ESO
    Telescopes at the La Silla Paranal Observatory under programme ID
    085.B-0567(A) and 088.B-0476(A). Based on observations made with
    the William Herschel Telescope operated on the island of La Palma
    by the Isaac Newton Group in the Spanish Observatorio del Roque de
    los Muchachos of the Instituto de Astrofísica de Canarias. }}

\author[A. J. Deason et al.]{A. J. Deason$^{1}$\thanks{E-mail:ajd75,vasily,nwe@ast.cam.ac.uk},
  V. Belokurov$^{1}$, N. W. Evans$^{1}$, S. E. Koposov$^{1,2}$, R. J. Cooke$^1$,
   \newauthor J. Pe\~{n}arrubia$^3$, C. F. P. Laporte$^{1,4}$, M. Fellhauer$^5$,
   M. G. Walker$^6$, E. W. Olszewski$^7$
\\ $^{1}$Institute of Astronomy, Madingley Rd, Cambridge, CB3 0HA,
\\ $^{2}$Sternberg Astronomical Institute, Moscow State University,
Universitetskiy pr. 13, Moscow 119991, Russia
\\$^{3}$ Ram\'on y Cajal Fellow, Instituto de Astrof\'isica de
Andalucia-CSIC, Glorieta de la Astronom\'ia s/n, 18008, Granada, Spain
\\ $^{4}$Max Planck Institute for Astrophysics, Karl-Schwarzschild
Strasse 1, 85740 Garching, Germany,
\\ $^{5}$Instituto de Astronom\'{i}a, Universidad de Concepci\'{o}n,
Casilla 160-C, Concpeci\'{o}n, Chile
\\ $^{6}$Hubble Fellow, Harvard Smithsonian Center for Astrophysics, 60 Garden
Street, Cambridge, MA 02138, USA
\\ $^{7}$Steward Observatory, University of Arizona, Tucson, AZ 85721,
USA}

\begin{document}

\date{June 2012}
\pagerange{\pageref{firstpage}--\pageref{lastpage}} \pubyear{2012}

\maketitle

\label{firstpage}

\begin{abstract}
We build a sample of distant ($D > 80$ kpc) stellar halo stars with
measured radial velocities. Faint ($20 < g <22$) candidate blue
horizontal branch (BHB) stars were selected using the deep, but wide,
multi-epoch Sloan Digital Sky Survey photometry. Follow-up
spectroscopy for these A-type stars was performed using the VLT-FORS2
instrument. We classify stars according to their Balmer line profiles,
and find 7 are bona fide BHB stars and 31 are blue stragglers
(BS). Owing to the magnitude range of our sample, even the
intrinsically fainter BS stars can reach out to $D \sim 90$ kpc. We
complement this sample of A-type stars with intrinsically brighter,
intermediate-age, asymptotic giant branch stars. A set of 4 distant
cool carbon stars is compiled from the literature and we perform
spectroscopic follow-up on a further 4 N-type carbon stars using the
WHT-ISIS instrument. Altogether, this provides us with the largest
sample to date of individual star tracers out to $r \sim 150$ kpc. We
find that the radial velocity dispersion of these tracers falls
rapidly at large distances and is surprisingly cold ($\sigma_r \approx
50-60$ km s$^{-1}$) between 100-150 kpc. Relating the measured radial
velocities to the mass of the Milky Way requires knowledge of the
(unknown) tracer density profile and anisotropy at these
distances. Nonetheless, by assuming the stellar halo stars between
$50-150$ kpc have a moderate density fall-off (with power-law slope
$\alpha < 5$) and are on radial orbits ($\sigma^2_t/\sigma^2_r < 1$),
we infer that the mass within 150 kpc is less than $10^{12}M_\odot$
and suggest it probably lies in the range $(5-10) \times
10^{11}M_\odot$. We discuss the implications of such a low mass for
the Milky Way.
\end{abstract}


\section{Introduction}

The formation and evolution of galaxies is fundamentally dependent on
their mass. We can measure the mass of a galaxy by a variety of
methods. The baryonic component is often inferred by assuming a
stellar initial mass function and converting the total integrated
starlight into mass. The overall mass, which is dominated by an unseen
dark matter component, requires more indirect methods such as
gravitational lensing, gas rotation curves, kinematic tracers or
timing arguments. Surprisingly, despite being a fundamental parameter,
the mass of our own Milky Way Galaxy is poorly known.

\begin{figure*}
  \centering
  \includegraphics[width=15cm, height=4cm]{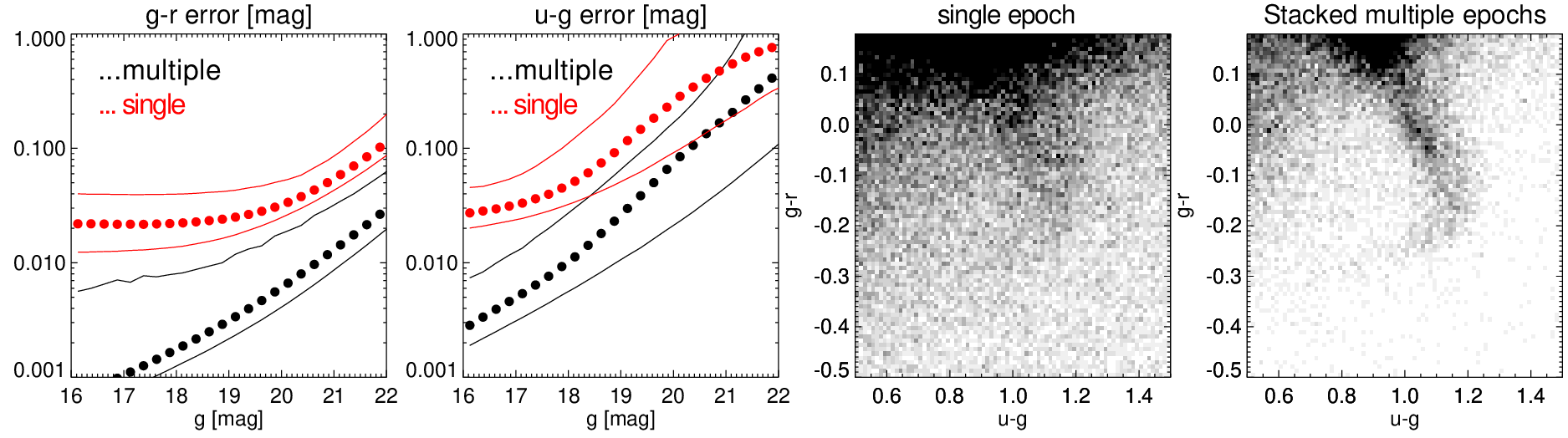}
  \caption[]{\small A comparison between single epoch and stacked
    multiple epoch SDSS stellar photometry. \textit{Left and Middle
    Left:} Median (dots), 5th and 95th percentiles (solid lines) of
    the PSF photometric error in the $g-r$ and $u-g$ colour of stars
    as a function of $g$ band magnitude. \textit{Middle Right and
    Right:} Density of star-like objects with $16<g<23$ in $u-g$,
    $g-r$ space. The BHB/BS `claw' is clearly visible in the stacked
    photometry.}
   \label{fig:overlaps}
\end{figure*}

The virial mass of the Milky Way has been measured (or inferred) to
lie within a large range of values, $5 \times 10^{11} <M_{\rm
vir}/M_\odot < 3 \times 10^{12}$. \cite{watkins10} found plausible
mass estimates in the range $7-34 \times 10^{11}M_\odot$ based on the
dynamics of satellite galaxies. This wide range of values partially
reflects the unknown properties of the tracer population, but is also
caused by a relatively low number of tracers together with uncertainty
as to whether some satellites are bound or unbound (e.g. Leo
I). Often, total mass estimates are extrapolations from the inner halo
to the virial radius. For example, \cite{xue08} inferred a virial mass
of $1 \times 10^{12}M_\odot$ based on the kinematics of blue
horizontal branch (BHB) stars out to 60 kpc. However, it is worth
remarking that such an extrapolation leads to a total mass that is not
wholly controlled by the data. Obviously, it is preferable to measure
the mass out to the virial radius rather than to infer it.

The mass within 50 kpc is constrained by the gas rotation curve
(e.g. \citealt{rohlfs88}), the orbit of the Magellanic stream
(e.g. \citealt{besla07}) and the dynamics of stellar halo stars
(e.g. \citealt{xue08}; \citealt{gnedin10}; \citealt{samurovic11}; \citealt{deason12}). Beyond
50 kpc, we rely on tracers such as satellite galaxies, globular
clusters and individual stars. The radial velocity dispersions of
these tracer populations provide a direct link to the dark matter,
which dominates the gravity field at such distances. In recent years,
several studies have shown that velocity dispersions in the Galactic
halo begin to decline at distances between $\sim 30-90$ kpc
(\citealt{battaglia05}; \citealt{brown10}), perhaps signifying that
the `edge' of the Milky Way is within reach. \cite{battaglia05} showed
that isothermal mass models predict much flatter
velocity dispersion profiles (with $\sigma_r \sim 100$ km s$^{-1}$)
than observed. However, despite recent
discoveries with Sloan Digital Sky Survey (SDSS) imaging
(e.g. \citealt{belokurov07}), there are unfortunately few known
satellite galaxies and globular clusters at distances beyond $80$ kpc,
which is where most of the mass uncertainty now arises. Thus, our
knowledge of the Milky Way's \textit{total} mass is limited by the
availability of dynamical tracers at large distances.

The best way of overcoming the paucity of objects is to use distant
halo stars as tracers. This has the advantage that there are many more
of them than satellite galaxies. Of halo stars, BHBs are
attractive targets because there exist well-understood methods for
their selection using colour-colour diagrams (e.g. \citealt{sirko04}),
as well as algorithms for removal of the principal contaminants like
blue stragglers (BS) (e.g. \citealt{clewley02}). The SDSS itself takes
advantage of the efficiency of the method and today has already
acquired spectra of thousands of BHBs over 1/4 of the sky
(\citealt{xue08}; \citealt{deason11a}; \citealt{xue11}). These stars
are however typically brighter than 19.5 magnitudes, and therefore at
best can reach 60 kpc (see e.g., \citealt{xue08}). Studies such as
those pioneered by \cite{clewley02} go deeper, but are limited to
small areas of the sky, and hence are susceptible to effects such as
substructure (\citealt{clewley05}). In this work, we go both deeper
and wider by making use of multi-epoch SDSS photometry to select distant BHB candidate stars.

We target distant BHB stars in the magnitude range $20 < g < 22$ (or a
distance range of $80 < D/\mathrm{kpc} < 200$ for BHBs) for
spectroscopic follow-up. These faint candidates require large 8-10m
class telescopes -- such as the European Southern Observatory (ESO)
Very Large Telescope (VLT) facility -- to obtain sufficient
signal-to-noise ratio (S/N) to classify the stars and measure their
radial velocities. We complement this sample of relatively old stars
with intermediate-age, asymptotic giant branch (AGB) stars. In
particular, we compile a sample of distant N type carbon (CN) stars,
which are bright ($M_{\rm R} \sim -3.5$) and can be detected out to
extremely large distances. In fact, their radial velocities can
comfortably be measured with 4m class telescopes such as the William
Herschel Telescope (WHT). In addition, unlike other AGB giant stars,
they can be cleanly selected using infrared photometry with little
contamination from dwarf stars (\citealt{totten98};
\citealt{totten00}). While these CN stars suffer less from photometric
contamination than BHB stars, they are intrinsically rare and many
(within $<100$ kpc) belong to the Sagittarius stream
(\citealt{ibata01}; \citealt{mauron04}). However, by constructing
distant halo samples of both populations we can ensure that our
results are not subject to a particular stellar type. 

The paper is arranged as follows. In Section 2, we outline our target
selection process and VLT-FORS2 spectroscopic follow-up programme for
candidate distant A-type stars. In addition, we outline the
classification of our targets as BHB or BS stars based on their Balmer
line profiles and assign distances according to this
classification. In Section 3, we compile a sample of distant CN type
stars in the literature with measured radial velocities. We perform
spectroscopic follow-up of a further four stars using the WHT-ISIS
instrument. We address the possibility that our distant stellar halo
stars belong to substructure in Section 4. The velocity dispersion
profile of our sample of distant stellar halo stars is analysed in
Section 5 and we consider the implications for the mass of the Milky
Way in Section 6, and sum up in Section 7.

\section{A-type stars}

We aim to measure radial velocities of distant ($D > 80$ kpc) BHB
stars. Selecting potential BHB candidates at such faint magnitudes
($20 < g < 22$) requires deep and accurate photometry. This is
especially important in the $u$ band, which gives the most
discriminating power between BHB and BS stars (\citealt{deason11b}). A
wide sky coverage is also beneficial to avoid targeting possible
substructures. Unfortunately, these two requirements are often
mutually exclusive with wider surveys having shallower photometry than
deeper pencil beam surveys. We have overcome this problem by making
use of the SDSS multi-epoch photometry. The deeper photometry provided
by stacking multi-epoch data allows us to target distant BHB stars
over a wide sky area. In the following sub-sections, we outline our
target selection and follow-up spectroscopic programme.

\subsection{Target Selection}
\label{sec:target_sel}

\begin{figure}
  \centering
  \includegraphics[width=8.5cm, height=5.5cm]{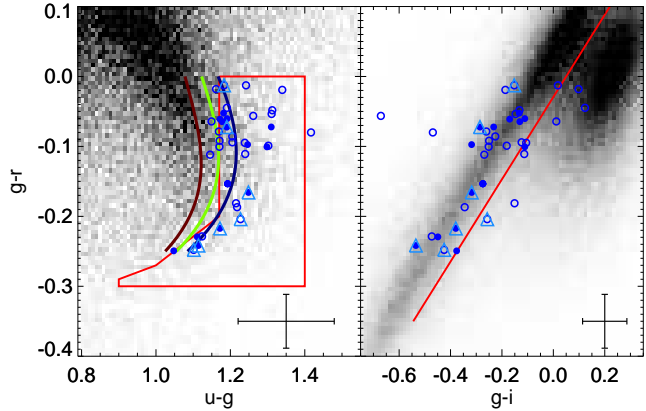}
  \caption[]{\small \textit{Left panel:} The density of stars with $19
    < g < 22$ in $u-g$, $g-r$ colour-colour space with stacked multi-epoch
    ($N_{\rm overlaps} > 2$) photometry. The red region shows the BHB
    star selection box. The BHB/BS star ridgelines
    (cf. \citealt{deason11b}) and approximate dividing line between
    the two populations are shown by the blue/red and green
    lines respectively. Stars selected for the P85 and P88 observations are shown
    by the open/filled blue symbols. The typical error of these points
    is shown in the bottom right corner. The light blue triangles
    around symbols indicate stars that are classified as BHBs in Section
    \ref{sec:class}. \textit{Right:} The density of stars in $g-r$,
    $g-i$ colour space. QSOs occupy the upper right hand region of
    this plot. The red line indicates the divide between QSOs and
    stars. Note that this cut was not applied in the P85
    observations.}
   \label{fig:sel}
\end{figure}

We have used deep photometry obtained by stacking multiple exposures
in the SDSS Stripe 82 to select faint BHB candidates with robust
colour measurements. Stripe 82 is a 2.5 degrees wide stripe that runs
at constant DEC $= 0$ deg and $-60 < \mathrm{RA}/\mathrm{deg} <
60$. During the SDSS Supernova search programme, locations along the
Stripe were imaged 20-50 times during the years 2000-2007. Individual
Stripe 82 images were stacked to produce deeper master frames, which
give a final catalogue of $u, g, r, i, z$ measurements for stars and
galaxies that is complete to 1-2 magnitudes fainter than that of SDSS
single-epoch observations. The left and middle left panels of
Fig. \ref{fig:overlaps} illustrate the improved precision in $g-r$
and $u-g$ for star-like objects in the stacked Stripe 82 data. The
middle right and right panel of Fig. \ref{fig:overlaps} show the
overall effect of obtaining deeper photometry around the BHB locus.

We have supplemented our BHB candidates selected from Stripe 82 with a
further sample chosen from the remaining overlaps in the SDSS 8th data
release sky coverage. About 40 per cent of the SDSS field of view is
observed more than once, as the survey stripes overlap. Typically, the
overlaps are observed at least twice, and occasionally 3-4
times. While shallower than Stripe 82 data, these multi-epoch data
improve the precision of magnitudes measured for faint candidates. We
select stellar objects with clean photometry, which have had 3 or more
detections within 0.5 arcseconds. For each object, we combine the
measurements from different observing runs and calculate average $u,
g, r$ magnitudes together with the scatter in each band (removing
those with standard deviations greater than 0.5 mag in each band). The
wider coverage of these SDSS overlaps allows us to complement the deep
photometry in Stripe 82 with a broader field of view.

This multi-epoch SDSS photometry allows us to easily identify A-type
stars which occupy the `claw' in $u-g$, $g-r$ colour space; the locus
of BHB stars is slightly redder in $u-g$ than BS stars
(\citealt{deason11b}). We define the following selection box to
maximise the number of BHB targets:
\begin{eqnarray}
\label{eq:col}
(u-g)_{\rm BHB}&=&[1.17, 1.17, 1.0, 0.9, 0.9, 1.4, 1.4, 1.17]  \\
(g-r)_{\rm BHB}&=&[0, -0.2, -0.27, -0.29, -0.3, -0.3, 0, 0] \nonumber
\end{eqnarray}
Stars in the magnitude range, $20 < g < 22$, that lie within this
colour-colour box, are identified as candidate distant BHB stars. Even
at brighter magnitudes, there is significant overlap between the BHB
and BS populations in $u-g$, $g-r$ colour-colour space and so we
expect there to be significant contamination by BS stars in our
selected targets. In particular, the relatively large errors in $u-g$
(see error bar in left panel of Fig. \ref{fig:sel}) means that there
can be significant scatter from BS stars into the BHB selection
region. However, we note that at such faint magnitudes ($g > 20$),
even BS stars can probe out to relatively large distances ($D \le 90$
kpc).

In Fig. \ref{fig:sel}, we show the targets selected for
spectroscopic follow-up. Stars observed in our first (P85) and second
(P88) observing runs are shown by the unfilled and filled circles
respectively. For the second run, we applied an extra cut in $g-r$,
$g-i$ space as shown in the right-hand panel of Fig. \ref{fig:sel}. By
excluding stars in the upper right region of this plot, we avoid
contamination from quasi-stellar objects (QSOs) in our sample.

\subsection{VLT-FORS2 follow-up spectroscopy}

We used the VLT FORS2 instrument (in service mode) to obtain low
resolution ($R \sim 800$) optical spectra of the 48 BHB
candidates. Observations were made in long-slit spectroscopy mode with
a $1.0$ arcsec slit. We used the 600B grism giving a dispersion of
$1.2$\AA\ per pixel. The spectral coverage, $\lambda=3400-6100$\AA\,
includes the Balmer lines H$\delta$, H$\gamma$ and
H$\beta$. Observations were taken in 2010 April-August (P85) and
2011-12 October-February (P88). For stars in the magnitude range $20 <
g < 22$, we required integration times between 20 minutes and 2 hours
to achieve a S/N ratio of 10 per resolution element.

The spectroscopic data were reduced using the \textsc{esorex} pipeline
provided by ESO. This pipeline performs all of the necessary steps
for reducing our science frames, including bias subtraction,
flat-fielding, spectral extraction and sky subtraction. A wavelength
calibration was applied with reference to HgCdHe arc lamp
observations. By comparing our wavelength solution to strong sky lines
(e.g. O\textsc{i} $\lambda 5577$\AA), we were able to make fine
corrections to the wavelength solution if necessary. The uncertainty
in the wavelength solution (typically $6$ km s$^{-1}$) is propagated
forward into our velocity errors. Finally, the data were
flux-calibrated with reference to standard star observations. 

From an inspection of the reduced spectra, we identified 4 targets
as QSOs and 1 target was too faint to be extracted. Thus, our sample
contains 43 A-type stars.

\subsection{BHB/BS Classification}
\label{sec:class}
\subsubsection{Template spectra}

\begin{figure*}
 \centering
 \includegraphics[width=16cm, height=10.67cm]{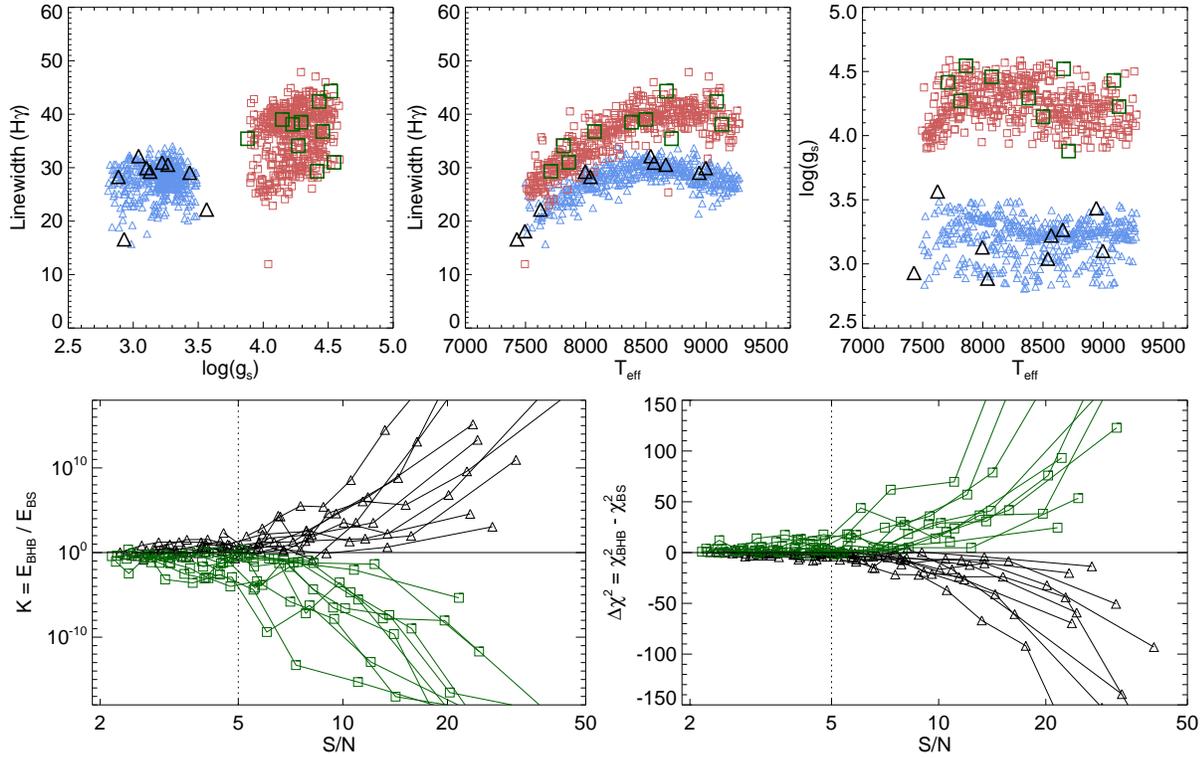}
 \caption{\small Top panels: Properties of our SDSS template
   spectra. We show the H$\gamma$ linewidths (similar trends are seen
   for H$\delta$ and H$\beta$), effective temperature and surface
   gravity. The blue triangles are the BHB templates and the red
   squares are the BS templates. We select 10 BHB and 10 BS test
   samples which are not in our template library. These are shown by
   the green squares (BS) and black triangles (BHB). Bottom panels: We
   degrade the S/N of the test samples and apply our template fitting
   routine. The left panel shows the Bayes factor (ratio of evidences)
   as a function of S/N. Stars with $K > 1$ are classified as BHB
   stars. Below S/N $< 5$ the stars become more difficult to
   classify. In the right hand panel, we show the difference between
   the minimum Chi-square values for the BHB and BS templates. BHB
   stars have negative $\Delta\chi^2$ values.}
 \label{fig:templates}
\end{figure*}

\begin{figure*}
 \centering
 \begin{minipage}{\linewidth}
   \centering
   \includegraphics[width=16cm, height=7cm]{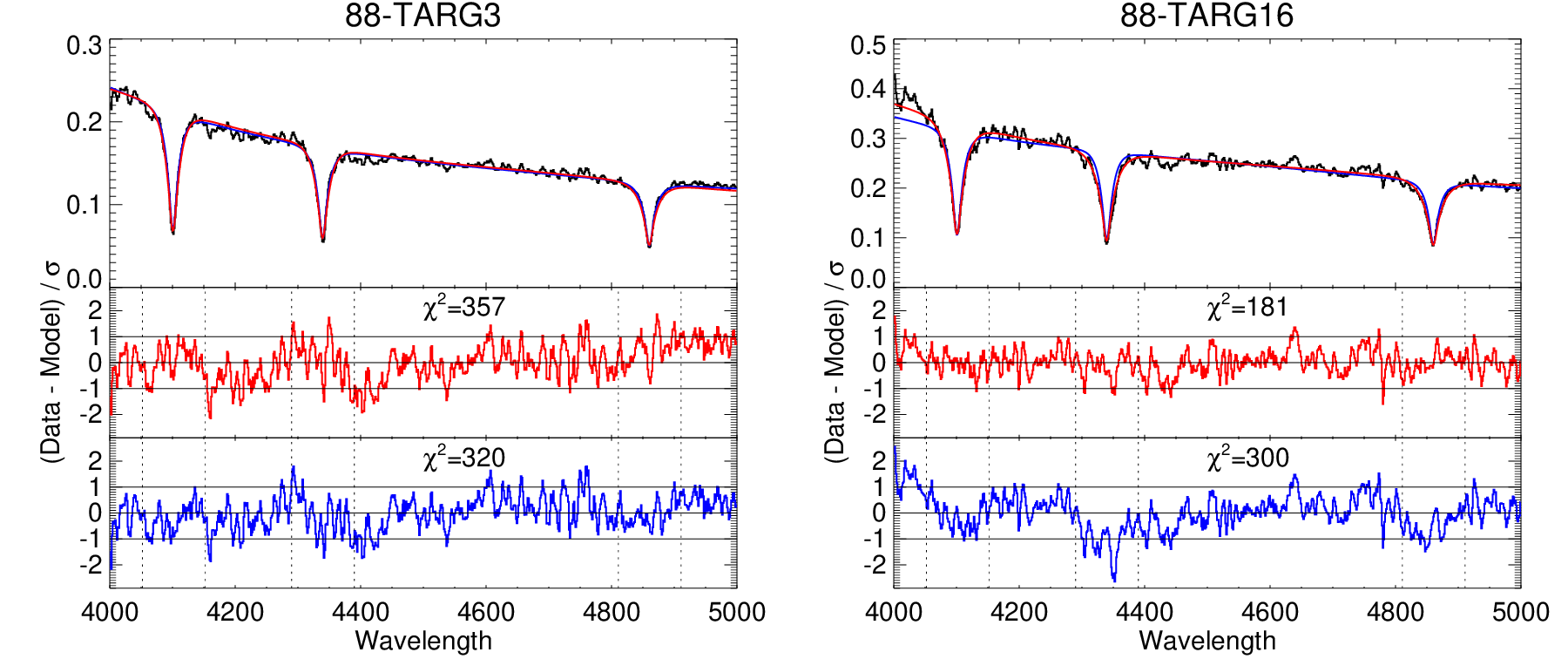}
 \end{minipage}
 \begin{minipage}{\linewidth}
   \centering
   \includegraphics[width=16cm, height=4cm]{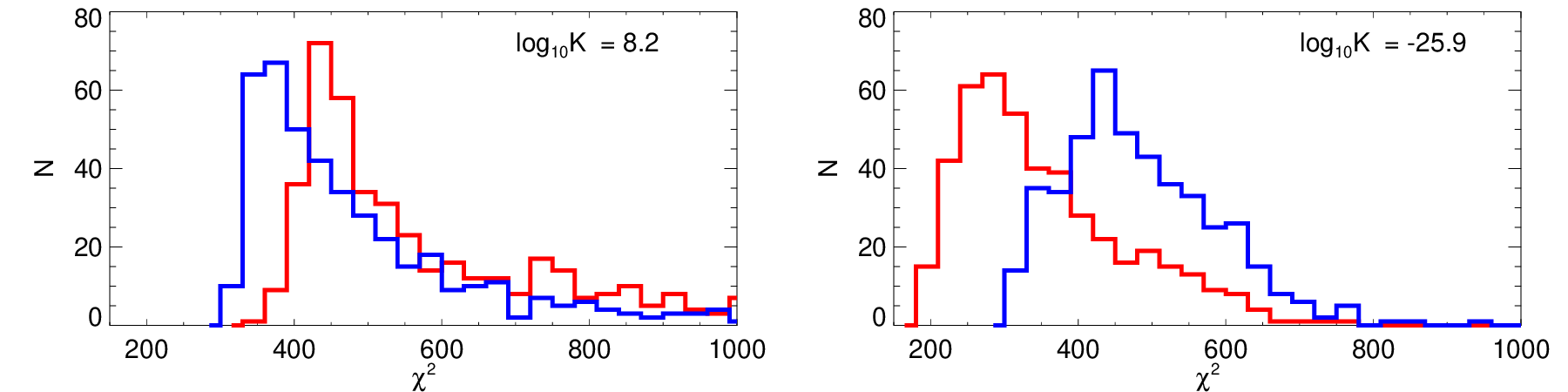}
 \end{minipage}
 \caption{\small Top panels: We show the minimum Chi-square template
   fits for two stars in our VLT-FORS2 sample. The residuals for the
   corresponding best-fit BHB (blue line) and BS (red line) templates
   are also shown. The dotted lines indicate the regions where the Chi
   square values are evaluated; these are centred on the three Balmer
   lines. Bottom panels: For the same two stars, we show the
   distribution of Chi-square values for all of the BHB and BS
   templates. The two stars (88-TARG3 and 88-TARG16) can be cleanly
   classified as a BHB or BS.}
 \label{fig:spectra}
\end{figure*}

To classify our A-type stars, we construct a sample of BHB
and BS template spectra using high S/N SDSS DR7 data. To select A-type
stars, we impose constraints on the (extinction corrected) $u-g$ and
$g-r$ colours, the surface gravity ($g_s$) and the effective
temperature ($T_{\mathrm{eff}}$), namely:

\begin{equation}
\begin{split}
0.7 < u-g < 1.4 \\
-0.3 < g-r < 0 \\
2.8 < \mathrm{log}(g_s) < 4.6 \\ 
7500 < T_{\mathrm{eff}}/ \mathrm{K} < 9300 
\end{split}
\end{equation}

At high S/N, BHB and BS stars can be classified according to their
surface gravity. The boundary between the two populations lies at
$\mathrm{log}(g_s)=3.5-4.0$. We select 1000 of these A-type stars with
high S/N ($>20$) that are approximately uniform in log$(g)$-$T_{\rm
eff}$ space and we exclude borderline cases ($3.5 <\mathrm{log}(g_s) <
4$, this excludes $<10$ per cent of each population). This sample is
roughly evenly split between BHB and BS stars. The top right hand
panel of Fig. \ref{fig:templates} shows the surface gravity and
effective temperature distribution of the BHB (blue triangles) and BS
(red squares) stars.

To construct templates from this sample, we parametrise the spectra in
the wavelength range $4000 < \lambda/${\AA}$ < 5000$. This spans the
wavelength coverage of our VLT-FORS2 sample and includes the Balmer
lines H$\delta$, H$\gamma$ and H$\beta$. We normalise the
spectra by fitting a polynomial of order 5 to the continuum flux
distribution, thus excluding regions that are affected by absorption
lines. We adopt the \cite{sersic68} function to fit the absorption line profiles of the
normalised spectra:
\begin{equation}
y=1.0-a\,\mathrm{exp}\left[-\left(\frac{|x-x_0|}{b}\right)^c\right]
\end{equation}
The parameters $x_0$ and $a$ give the wavelength and line depth at
the line centre respectively. The parameter $b$ provides a measure of
the linewidth and the parameter $c$ quantifies the line shape. The
model profiles are convolved with the full-width at half-maximum
(FWHM) resolution ($\sim 2.3$ \AA) of the SDSS spectra. Our profiles are fitted using the
publicly available IDL MPFIT\footnote{http://purl.com/net/mpfit}
programme (\citealt{mpfit}).

Thus, each A-type `template' star is parametrised by the (5th order
polynomial) continuum and the Sersic profiles of the H$\delta$,
H$\gamma$ and H$\beta$ Balmer lines. This parametrisation is used to
produce noise-free templates. The Sersic profile parameters and
continuum shape of the stars are governed by their surface gravity and
effective temperature. Thus, when fitting the data, we vary the
template ID, which is equivalent to varying log$(g_s)$ and $T_{\rm
eff}$. In the top panels of Fig. \ref{fig:templates}, we show the
H$\gamma$ linewidths, effective temperature and surface gravity for
our template stars.

Each of these template stars is a model to which we can compare our
VLT-FORS2 data. The models are convolved with the FWHM resolution
applicable to our VLT-FORS2 observations ($\sim 4$ \AA), where we
measure the FWHM from the arc lines. For each model, there are two
free parameters; the velocity and a constant scale factor. \textbf{The
  derived velocities are based solely on the three strong Balmer
  lines, and both BHB and BS template fits give very similar
  velocities}. The
Chi-square value of each fit can then be used to decide if BHB star
templates or BS star templates provide a better description of the
data. To statistically determine the preferred model type, we use the
ratio of evidences (or the Bayes factor):

\begin{eqnarray}
\label{eqn:evidence}
K&=&\frac{E_{\rm BHB}}{E_{\rm BS}} =\frac{\int
  \mathrm{Prior(\theta_{\rm BHB})} \, \mathrm{exp}\left(-\chi^2(\theta_{\rm
    BHB})/2\right) \, \mathrm{d}\theta_{\rm BHB}}{\int \mathrm{Prior(\theta_{\rm BS})}
  \, \mathrm{exp}\left(-\chi^2(\theta_{\rm BS})/2\right) \,
  \mathrm{d}\theta_{\rm BS}} \\ \nonumber
&\simeq &\frac{\sum_i    \, \mathrm{Prior(\psi)}_{\rm BHB,
      i} \, \sqrt{|\mathrm{Cov(\psi)}|_{\rm BHB,i}} \,
  \mathrm{exp}\left(-\chi^2_{\rm BHB,i}/2\right)}{\sum_i   \,
  \mathrm{Prior(\psi)}_{\rm BS,  i} \, \sqrt{|\mathrm{Cov
      (\psi)}|_{\rm BS,i}} \,
  \mathrm{exp}\left(-\chi^2_{\rm BS,i}/2\right)}
\end{eqnarray}

Here, $\theta$ denotes the parameter space over which the template
models span; the template IDs (which cover a range of surface
gravities and effective temperatures) and the free parameters in each
template fit. The sum is performed over all
templates, where $\psi$ denotes the two free
parameters for each template fit, $v_{\rm los}$ and a constant
scale factor. A Gaussian approximation is assumed for the
marginalisation over the $\psi$ parameters, which then depends on the
corresponding covariance matrix ($|\mathrm{Cov}(\psi)|$, see e.g. \citealt{mackay03}). We assume uniform
priors for all parameters except the constant scale factor, for
which we use a Jeffrey's scale invariant prior (\citealt{jeffreys61}).

Equation \ref{eqn:evidence} gives the ratio of marginalised
likelihoods. However, we can compute the \textit{posterior} odds by
multiplying the likelihood ratio by the prior odds. In our case, the
prior odds could be the number ratio of BHB-to-BS stars,
i.e. $\tilde{K}= \left(f_{\rm BHB}/f_{\rm BS}\right)\, K$. In most
cases, we do not know the number ratio of BHB-to-BS stars prior and by
assuming $f_{\rm BHB}=f_{\rm BS}$, the prior evidence ratio reduces to
$\tilde{K}=K$.

We test our method of classifying BHB and BS stars by randomly
selecting 10 BHB and 10 BS stars from the SDSS sample. These stars are
not included in our template spectra sample. They are shown by the
black triangles (BHB) and green squares (BS) in
Fig. \ref{fig:templates}. We perform our fitting routine on these test
cases and vary the S/N by degrading the spectra. The results of this
exercise are illustrated in the bottom panels of
Fig. \ref{fig:templates}. In the left panel, we show the Bayes
factor, $K$, as a function of S/N. The right panel shows the
difference in Chi-square values between the best fitting BHB and BS
models. The BHB stars should have $K > 1$ and $\Delta\chi^2 < 0$. At
S/N $>5$, the classification is almost always correct ($> 90\%$) and
stars can easily be classified when S/N $>10$. However, we find that
below S/N $<5$, the stars become more difficult to classify.

We observed three `standard' BHB stars over our observing run. These
are bright ($V \sim 15$) stars classified as BHBs in the
literature. Two of these stars were observed in P85 and reside in the
M5 globular cluster (cf. \citealt{clewley02}) and one is taken from
the \cite{xue08} sample of SDSS BHB stars and was observed in P88. We
applied the same data reduction procedures to these standard stars as
our main sample. We apply our template fitting technique to these
stars and obtain very high evidence ratios ($\mathrm{log}_{10}K >
100$), as expected for such high S/N spectra (S/N $\sim 100$).

\subsubsection{Results}

\begin{table*}
\centering
\renewcommand{\tabcolsep}{0.2cm}
\renewcommand{\arraystretch}{1.5}
\begin{tabular}{| c c c c c c c c c c c c c c c |}
\hline
ID & ra  & dec & $g_{\mathrm{mag}}$ & (S/N) & $\chi^2_{\rm BHB}$ & $\chi^2_{\rm BS}$ & $\mathrm{log}_{10}K$ & $\mathrm{log}_{10}\tilde{K}$ & $T_{\rm eff}$ & ${\rm log} (g_s)$ & $V_{\mathrm{hel}}$  & D & Class\\
&  [deg] & [deg] &  & $\mathrm{pix}^{-1}$ & & & & & K & & km s$^{-1}$  & kpc &\\
\hline
85-TARG2     & 219.2258 & 0.8461 & 21.8 & 6.4 & 262.7 & 235.9 & -5.87 & -6.60 & 8529 & 4.31 & -38 $\pm$ 23 & 72 $\pm$ 17 & BS \\
85-TARG3     & 220.6565 & -1.1907 & 21.7 & 6.3 & 268.9 & 254.4 & -3.17& -3.90 & 8286 & 4.32 & 86 $\pm$ 21 & 60 $\pm$ 14 & BS \\
85-TARG4     & 202.4456 & 8.3195 & 21.5 & 10.1 & 311.3 & 305.4 & -1.25 & -1.98 & 7568 & 4.04 & 8 $\pm$ 13 & 55 $\pm$ 13 & BS \\
85-TARG5     & 217.5596 & 7.3932 & 21.5 & 9.8 & 338.3 & 323.3 & -3.32 & -4.05 & 7614 & 4.08 & -73 $\pm$ 20 & 59 $\pm$ 14 & BS \\
85-TARG7     & 204.0844 & 4.7730 & 21.5 & 10.3 & 405.6 & 397.0 & -1.84& -2.57 & 7578 & 4.05 & 64 $\pm$ 21 & 62 $\pm$ 14 & BS \\
85-TARG10    & 204.9583 & 3.3285 & 21.1 & 8.2 & 348.6 & 341.7 & -2.06 & -2.78 & 7851 & 4.52 & 93 $\pm$ 28 & 54 $\pm$ 13 & BS \\
85-TARG12    & 226.0665 & 3.1375 & 21.0 & 10.5 & 281.4 & 266.7 & -3.25 & -3.97 & 8286 & 4.32 & 6 $\pm$ 20 & 54 $\pm$ 12 & BS \\
85-TARG13    & 233.7010 & -0.8291 & 21.0 & 11.0 & 363.8 & 300.0 & -13.92 & -14.65 & 9182 & 4.48 & 116 $\pm$ 18 & 48 $\pm$ 11 & BS \\
85-TARG15    & 151.2014 & 11.7011 & 21.2 & 4.9 & 199.6 & 191.9 & -1.79 & -2.52 & 7614 & 4.08 & 67 $\pm$ 28 & 71 $\pm$ 16 & BS \\
85-TARG17    & 192.8952 & 6.2015 & 20.9 & 7.6 & 255.1 & 247.7 & -1.94 & -2.67 & 8286 & 4.32 & 278 $\pm$ 24 & 46 $\pm$ 11 & BS \\
85-TARG18    & 200.6029 & 10.8438 & 21.2 & 9.1 & 341.4 & 330.3 & -1.94 & -2.66 & 7614 & 4.08 & 137 $\pm$ 37 & 53 $\pm$ 12 & BS \\
85-TARG20    & 227.1404 & -0.8308 & 20.5 & 8.7 & 370.5 & 299.7 & -15.45 & -16.18 & 8013 & 4.27 & 142 $\pm$ 36 & 40 $\pm$ 9 & BS \\
85-TARG21    & 203.6062 & 3.4247 & 20.8 & 6.7 & 221.2 & 209.4 & -2.95 & -3.67 & 8675 & 4.18 & 15 $\pm$ 28 & 50 $\pm$ 12 & BS \\
85-TARG23    & 138.2685 & 2.7640 & 20.8 & 7.6 & 274.4 & 259.2 & -3.71 & -4.44 & 8499 & 4.14 & 305 $\pm$ 36 & 48 $\pm$ 11 & BS \\
85-TARG24    & 203.2243 & 5.1525 & 20.5 & 4.1 & 442.7 & 260.2 & -39.81 & -40.53 & 8437 & 4.44 & 25 $\pm$ 13 & 55 $\pm$ 13 & BS \\
85-TARG33    & 206.9920 & -1.1284 & 20.3 & 16.9 & 279.1 & 285.4 & 2.17 & 1.44 & 8406 & 3.27 & 2 $\pm$ 20 & 83 $\pm$ 4 & BHB \\
85-TARG34    & 133.3904 & -0.3267 & 20.3 & 6.3 & 213.5 & 221.9 & 2.54 & 1.81 & 8477 & 3.21 & 131 $\pm$ 18 & 88 $\pm$ 4 & BHB \\
85-TARG36    & 204.1527 & 5.6618 & 20.3 & 3.8 & 350.7 & 331.6 & -4.19 & -4.92 & 7614 & 4.08 & 34 $\pm$ 19 & 32 $\pm$ 7 & BS \\
85-TARG38    & 179.1346 & 6.6239 & 20.3 & 4.7 & 354.9 & 280.8 & -16.17 & -16.90 & 8494 & 4.47 & 37 $\pm$ 16 & 39 $\pm$ 9 & BS \\
88-TARG2     & 45.0211 & -0.3606 & 20.8 & 14.8 & 356.6 & 299.1 & -12.65 & -13.38 & 8708 & 4.06 & 164 $\pm$ 13 & 65 $\pm$ 15 & BS \\
88-TARG3     & 28.1598 & 0.0305 & 21.0 & 12.1 & 320.1 & 357.1 & 8.16 &7.43 & 8700 & 3.07 & -73 $\pm$ 10 & 119 $\pm$ 6 & BHB \\
88-TARG4     & 23.7335 & 0.9832 & 21.0 & 14.6 & 414.6 & 353.2 & -13.49& -14.22 & 8442 & 4.47 & -181 $\pm$ 15 & 52 $\pm$ 12 & BS \\
88-TARG5     & 27.1481 & 0.6968 & 21.3 & 18.7 & 348.4 & 315.5 & -7.25 & -7.97 & 9074 & 4.36 & -11 $\pm$ 38 & 86 $\pm$ 20& BS \\
88-TARG7     & 12.6871 & 13.9173 & 21.0 & 14.7 & 283.6 & 292.9 & 2.32 & 1.59 & 8801 & 3.28 & -121 $\pm$ 13 & 116 $\pm$ 5 & BHB \\
88-TARG8     & 22.4696 & 3.2119 & 21.1 & 11.5 & 282.0 & 289.9 & 1.68 & 0.96 & 8795 & 3.32 & -47 $\pm$ 14 & 133 $\pm$ 6 & BHB \\
88-TARG9     & 10.3506 & 0.1664 & 20.8 & 16.0 & 334.1 & 333.1 & -0.47& -1.20 & 8529 & 4.31 & -129 $\pm$ 35 & 54 $\pm$ 12 & BS(?) \\
88-TARG15    & 42.0548 & -6.8055 & 20.6 & 10.8 & 351.8 & 278.7 & -15.89 & -16.62 & 8442 & 4.47 & -52 $\pm$ 15 & 43 $\pm$ 10 & BS \\
88-TARG16    & 65.2176 & -0.6718 & 20.2 & 10.5 & 300.2 & 181.2 & -25.86 & -26.59 & 9074 & 4.36 & -69 $\pm$ 17 & 44 $\pm$ 10 & BS \\
88-TARG32    & 16.4854 & -1.2477 & 21.2 & 16.9 & 351.4 & 272.7 & -17.15 & -17.87 & 8442 & 4.47 & -11 $\pm$ 19 & 56 $\pm$ 13 & BS \\
88-TARG33    & 50.1435 & -0.7992 & 21.0 & 14.4 & 381.7 & 312.2 & -15.11 & -15.84 & 8442 & 4.47 & -100 $\pm$ 16 & 49 $\pm$ 11 & BS \\
88-TARG34    & 4.4821 & 0.0631 & 21.3 & 12.8 & 165.3 & 172.5 & 1.41 & 0.68 & 8233 & 3.24 & -88 $\pm$ 16 & 151 $\pm$ 7 & BHB \\
88-TARG35    & 23.7335 & 0.9831 & 21.0 & 16.5 & 361.9 & 234.4 & -27.88 & -28.61 & 8297 & 4.45 & -173 $\pm$ 12 & 48 $\pm$ 11 & BS \\
88-TARG36    & 20.7336 & -0.6500 & 21.2 & 20.5 & 298.3 & 270.6 & -6.38 & -7.11 & 8175 & 4.35 & -36 $\pm$ 14 & 53 $\pm$ 12 & BS \\
88-TARG37    & 1.5332 & -0.2538 & 21.3 & 11.8 & 333.6 & 291.6 & -9.13 & -9.86 & 8297 & 4.45 & -363 $\pm$ 19 & 54 $\pm$ 12 & BS \\
88-TARG41    & 41.3665 & -9.0043 & 20.8 & 11.6 & 210.1 & 186.3 & -5.27 & -6.00 & 9182 & 4.48 & -78 $\pm$ 26 & 49 $\pm$ 11 & BS \\
88-TARG100   & 49.0847 & -0.1127 & 20.8 & 11.6 & 269.9 & 272.5 & 0.50 & -0.23 & 7522 & 3.37 & -35 $\pm$ 13 & 117 $\pm$ 5 & BHB(?) \\
88-TARG105   & 40.9579 & -0.6469 & 20.2 & 14.1 & 197.4 & 192.6 & -1.36 & -2.09 & 7959 & 4.23 & 34 $\pm$ 18 & 35 $\pm$ 8 & BS \\
88-TARG110   & 49.0308 & -0.5904 & 20.0 & 14.2 & 363.5 & 310.1 & -11.77 & -12.49 & 8297 & 4.45 & -75 $\pm$ 13 & 33 $\pm$ 8 & BS \\
\hline
 \end{tabular}
\caption[]{\small We list the classifiable A-type stars observed in
  our VLT run. The columns list the ID, the right ascension and
  declination, the (extinction corrected) magnitude, the S/N per
  resolution element, the \textit{minimum} Chi-square value for the
  BHB and BS star templates, the ratio of the BHB and BS evidences
  ($K$), the evidence ratio weighted by the fraction of BHB-to-BS star
  priors ($\tilde{K}$), the effective temperature and surface gravity
  of the best fitting template model, the observed heliocentric
  velocities, and the distance derived from the $g-r$ colours
  according to the stellar classification (i.e. BHB or BS). The final
  column gives the assigned classification with a question mark
  indicating borderline cases.}
\label{tab:results}
\end{table*}

A small number (5) of our VLT-FORS sample are too noisy to classify
(S/N $<5$) and so we apply our fitting method to the remaining (38)
classifiable A-type stars. The results are summarised in Table
\ref{tab:results}, where we give the measured heliocentric velocity
and final classification for each star. We show an example of a BHB
(88-TARG3, left column) and a BS (88-TARG16, right column) star in
Fig. \ref{fig:spectra}. The top panels show the spectrum (black line)
and the best fitting BHB (blue line) and BS (red line) models
overplotted. The residuals for each fit are also shown. The bottom
panels show the distribution of $\chi^2$ values for each template
fit. In these two cases, the stars are easily identified as a BHB or
BS.

We classify stars with $K > 1$ as BHB stars. This classification
method identifies 7 BHB stars and 31 BS stars. \cite{jeffreys61}
define $K > 3$ or $K < 1/3$ as substantial evidence for (or against) a
particular model. Following this interpretation, two of our stars
(88-TARG100 and 88-TARG9) are identified as borderline cases. While we
do not have prior knowledge of the fraction of BHB and BS stars, we
can iteratively compute $\tilde{K}$. We begin by assuming $f_{\rm
BHB}=0.5$ and then iterate until the fraction of BHB stars
converges (to $f_{\rm BHB}=0.16$). The low fraction of BHB stars means that $\tilde{K} < K$
and in one borderline case (88-TARG100) $K > 1$, but including the
prior factor gives $\tilde{K} < 1$.

\subsection{Distances}

BHB stars are near ideal `standard candles' owing to their intrinsic
brightness and narrow range of absolute magnitudes. However, BS stars
are fainter (by $\sim 2$ mags) and have a much larger range of
intrinsic luminosities. We assign absolute magnitudes to the two
populations using the absolute magnitude-colour relations given by
equation 7 in \cite{deason11b}. The spread in absolute magnitude about
this relation is fairly narrow for BHBs ($\Delta M_g \sim 0.15$) as
compared to BS stars ($\Delta M_g \sim 0.5$). Heliocentric distances
are assigned to our A-type stars according to their classification as
a BHB or BS star. The BS stars span a distance range $30 <
D/\mathrm{kpc} < 90$, whilst the BHB stars probe much further distances
$80 < D/\mathrm{kpc} < 150$.

\section{Cool Carbon stars}

We complement our sample of A-type stars with distant ($D > 80$ kpc)
cool carbon stars, which have radial velocities published in the
literature. In particular, we focus on N-type carbon (CN) stars owing
to their clean photometric selection using $JHK$ bands and their
bright intrinsic luminosity (see e.g. \citealt{totten98}); CN stars with
apparent magnitudes as bright as $r \sim 15$ can reach out to $D \sim
80-100$ kpc. In addition to the literature sample of CN stars (with details
provided in table \ref{tab:carbon}), we obtain follow-up spectroscopy for four CN stars
which do not have radial velocity measurements. We outline their
selection and follow-up spectroscopy below.

\begin{figure}
  \centering
  \includegraphics[width=8.5cm, height=7.5cm]{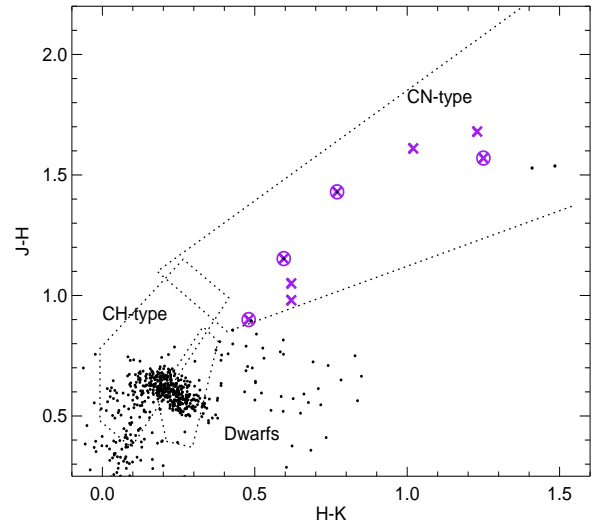}
  \caption[]{\small $J-H, H-K$ colours of very red ($g-r > 1.75$) SDSS
  stars with small proper motion ($\mu < 9$ mas yr$^{-1}$)
  cross-matched with UKIDSS stars. The classification boundaries for
  CH, CN and dwarf stars are taken from \cite{totten00}. The purple
  crosses show the distant CN stars given in Table
  \ref{tab:carbon}. The circled indicate stars for which we perform
  follow-up spectroscopy.}
   \label{fig:carbon_jhk}
\end{figure}

We target CN stars using SDSS optical photometry and Two Micron All
Sky Survey (2MASS) or UKIRT Infrared Deep Sky Survey (UKIDSS) infrared
photometry. We select SDSS stars with very red colours ($g-r > 1.75$)
and with no appreciable proper motion ($\mu < 9$ mas yr$^{-1}$). These
stars are then cross-matched with the infrared catalogues to obtain
their $JHK$ magnitudes. In Fig. \ref{fig:carbon_jhk}, we show the
resulting candidates from UKIDSS photometry by the black dots. The
CN-type, CH-type and dwarf star classification boundaries are shown by
the dotted lines (see \citealt{totten00}). The CH-type stars and
dwarfs stars greatly outnumber the CN type stars. This is unsurprising
given that the CN type stars are intrinsically rare
(\citealt{totten98}) and also probe much greater distances. The purple
crosses show the distant CN stars given in Table \ref{tab:carbon}. The
circles indicate stars without radial velocity measurements for which
we perform follow-up spectroscopy. Two of these stars are selected
from cross-matching SDSS with either 2MASS or UKIDSS photometry. The remaining
two are classified as CN stars by \cite{mauron08} and have estimated
distances $D> 100$ kpc.

\subsection{WHT-ISIS follow-up Spectroscopy}

We used the WHT ISIS instrument to obtain high resolution ($R \sim
5000$) optical spectra of the 4 CN type stars. Long slit spectroscopic
observations were made in good conditions (seeing $\sim 1.0$ arcsec) with
a $0.8$ arcsec wide slit. We used the R1200 grating giving a
dispersion of $0.26$\AA\ per pixel. The spectral coverage, $\lambda
\simeq 7500-8500$\AA, includes several CN absorption bands. These
observations were taken on 2012 March 27-31. For stars in the
magnitude range $15 < r < 19$, we required integration times between
300 seconds and 1 hour to achieve a S/N ratio of 10 per resolution
element.  We follow the data reduction procedure outlined in
\cite{totten98}. In short, the spectroscopic data were reduced using
the standard \textsc{iraf} software packages. First, the science
frames were bias subtracted and flat-fielded. Background regions were
selected for sky subtraction during the extraction of the science
data. The extracted stellar spectra were wavelength calibrated with
reference to CuNe+CuAr arc lamp observations. As a final step, the
reduced data were flux calibrated with observations of flux standard
stars.

We also observed several carbon star radial velocity standards. These
stars were observed and reduced with identical instrument setup and
data reduction procedures as described above for our science
targets. We use these standard stars as templates to determine the
velocities of our science targets using cross-correlation
techniques. With the high resolution and modest signal-to-noise of the
data (S/N $\sim 10$), the near-infrared CN absorption bands are well
defined and lead to good-quality cross-correlation peaks. There are
several velocity errors that affect the final results including
wavelength calibration uncertainties, mismatched template versus
science target stars and random errors due to photon counting. The
combined effect of all these errors gives velocity uncertainties of
$\sim 10$ km s$^{-1}$.
\begin{table*}
\centering
\renewcommand{\tabcolsep}{0.3cm}
\renewcommand{\arraystretch}{1.5}
\begin{tabular}{| c c c c c c c c c c |}
\hline
ID & RA (J2000) & DEC (J2000) & $J_{\mathrm{mag}}$ &
$H_{\mathrm{mag}}$ & $K_{\mathrm{mag}}$ & $V_{\mathrm{hel}}$ & $D_{
  \rm M}$
& $D_{\rm TI}$ & Ref. \\
& [deg] & [deg] &   &  &  & km s$^{-1}$ & kpc & kpc & \\
\hline
B1429-0518 & 218.1198 & -5.5216 & 13.9 & 12.3 & 11.0 & 77 $\pm$ 6 & 70 & 94 & TI98/00 \\
B1450-1300 & 223.4304 & -13.2180 & 13.9 & 12.3 & 11.3 & 120 $\pm$ 5 & 76 & 101 & TI98/00\\
J2246-2726 & 341.6206 & -27.4501 & 14.9 & 13.9 & 13.3 & 2 $\pm$ 10 & 130 & 160 & M05\\
J1141-3341 & 175.4258 & -33.6928 & 13.9 & 12.8 & 12.2 & 144 $\pm$ 10 & 82 & 102 & M05\\
J1446-0055 & 221.6295 & -0.9168 & 13.6 & 12.2 & 11.4 & 15 $\pm$ 10 & 71 & 95 & UKIDSS-ISIS\\
J1301+0029 & 195.3269 & 0.4975 & 14.9 & 13.7 & 13.1 & 61 $\pm$ 10 & 153 & 169 & UKIDSS/M08-ISIS\\
J1725+0300 & 261.4764 & 3.0072 & 14.8 & 13.9 & 13.4 & -72 $\pm$ 10 & 115 & 136 & M08-ISIS\\
J0905+2025 & 136.4432 & 20.4106 & 15.1 & 13.6 & 12.3 & 200 $\pm$ 10 & 140 & 165 & 2MASS-ISIS\\
 \hline
\end{tabular}
\caption[]{We list the distant CN stars with radial velocities from
the literature or velocities measured in this work. The columns give
the ID, the right ascension and declination, the infrared $J, H$ and
$K$ magnitudes, the observed heliocentric velocities and the estimated
distances using the \cite{mauron04} prescription or the
\cite{totten00} relation. We adopt the mean of these distances with an
associated distance error of 25 per cent. The final column lists the
reference for the photometry and measured velocities:
\citealt{mauron05} (M05); \citealt{mauron08} (M08); \citealt{totten98}
(TI98); \citealt{totten00} (TI00). }
\label{tab:carbon}
\end{table*}

\begin{figure}
  \centering
  \includegraphics[width=8.5cm, height=7.5cm]{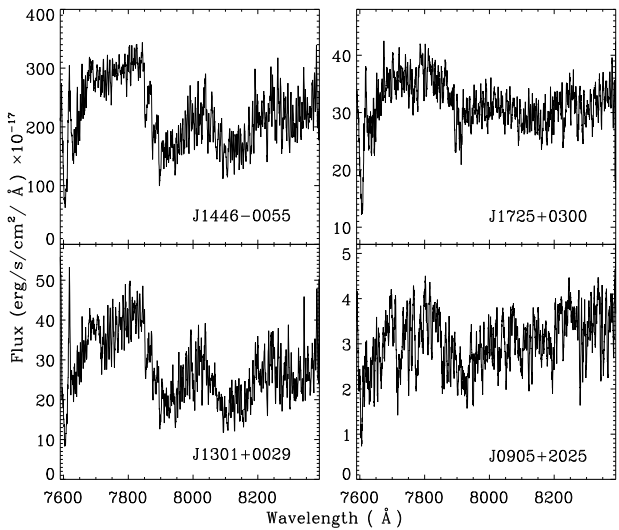}
  \caption[]{\small High resolution spectra for the four CN-type
    stars. We concentrate on the wavelength region $7500 <
    \lambda$/\AA$ < 8500$ where there are several CN absorption
    bands. The apparent `noisiness' of the data is due to the many CN
    absorption lines, and not entirely due to photon counting errors.}
   \label{fig:carbon_spec}
\end{figure}

\subsection{Distances}
We assign distances to the carbon stars using their $K$ band
magnitude. We use infrared rather than optical photometry, since the
former is more robust to extinction and variability
effects. \cite{totten00} estimated distances to CN type carbon stars
by calibrating against Galactic satellites. This calibration is
largely dominated by stars belonging to the Large Magellanic Cloud
(LMC) and Small Magellanic Cloud (SMC). \cite{mauron04} use carbon
stars in Sagittarius (Sgr) to assign distances to their stars. The
authors note that the carbon stars of Sgr are less luminous by $\sim
0.5$ mag in the K band than the Carbon stars in the LMC. Thus, the
distances derived by \cite{mauron04} are $\sim 25$ percent closer than
those derived by \cite{totten00}. In Table \ref{tab:carbon}, we give
both the \cite{totten00} ($D_{\rm TI}$) and \cite{mauron04} ($D_{\rm
M}$) distances of our CN type stars. We adopt the mean of these two
distances in this work and assume errors of $\sigma_D \sim 25$
percent.

\medskip
\noindent
Our final sample consists of 8 CN type stars in the distance range $80
< D/\mathrm{kpc} < 160$, 7 BHB stars with $80 < D/\mathrm{kpc} < 150$
and 31 BS stars with $30 < D/\mathrm{kpc} < 90$. This is
the largest sample of distant stellar halo stars with measured radial
velocities to date.

\section{Field stars or substructure?}
\begin{figure*}
  \centering
  \includegraphics[width=15cm, height=9.5cm]{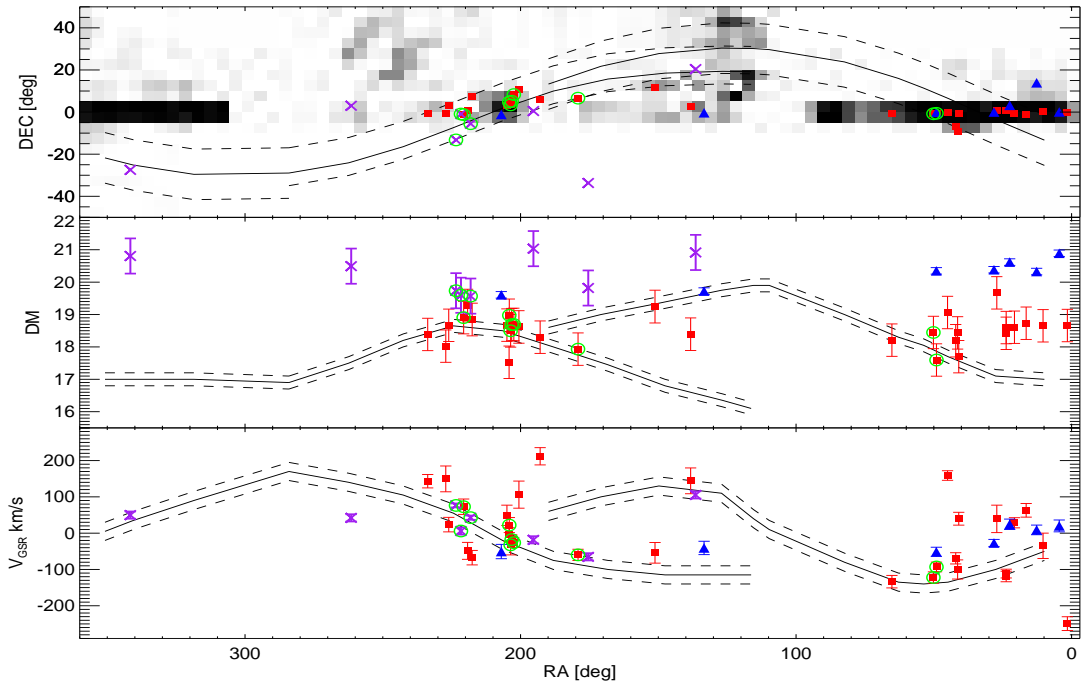}
  \vspace{-10pt}
  \caption[]{\small \textit{Top panel:} The distribution of our A-type
    and CN-type stars on the sky. The red squares, blue triangles and
    purple crosses indicate BS, BHB and CN stars respectively. The
    shaded regions show the density of stars which have multi-epoch
    photometry ($N_{\rm overlap} > 3$). \textit{Middle panel:}
    Distance modulus as a function of right ascension. \textit{Bottom
    panel:} Galactocentric velocity as a function of right
    ascension. The black lines in all three panels show the
    approximate tracks for Sgr leading and trailing stream
    stars. Green circles indicate stars that lie inside all three
    regions (declination, distance and velocity) appropriate for Sgr
    stars. We find 8 BS stars associated with Sgr and one possible BHB
    star (if its classification as a BHB is incorrect). Three CN stars
    at RA $\sim 220$ deg are likely associated with the Sgr leading
    arm. The remaining distant BHB and CN stars show no evidence for
    spatial or velocity clustering and we consider it unlikely that
    they all belong to a common substructure.}
     \label{fig:sgr}
\end{figure*}

\begin{figure}
  \centering
  \includegraphics[width=8.5cm, height=7cm]{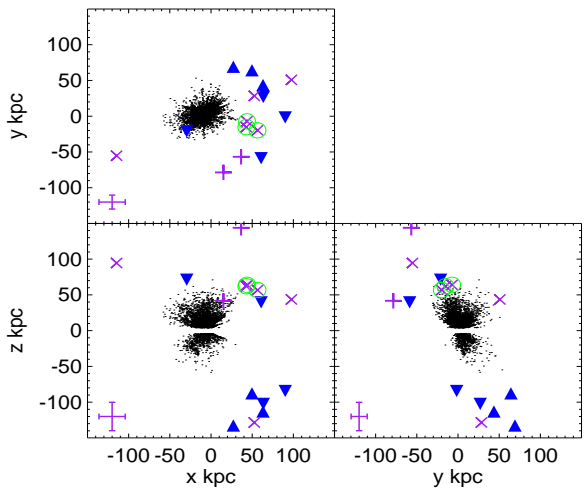}
   \vspace{-20pt}
  \caption[]{\small The $x,y$ and $z$ positions of our distant BHB
     (blue triangles) and
     CN (purple crosses) stellar halo stars. The stars with positive
     radial velocities ($v_{\rm GSR} > 0$) are indicated by blue triangles or purple crosses, while the stars with negative radial velocities ($v_{\rm GSR} < 0$) are indicated by the blue upside down triangles and the purple plus symbols. For comparison, we show the spatial distribution of the \cite{xue11} sample of BHB stars selected from SDSS. The green circles indicate the 3 CN stars that may belong to the Sgr leading arm. The uncertainty in the CN star coordinates
     is shown by the error bars in the bottom-left corner of every
     panel. The distance errors for the BHB stars are smaller than the
     symbol sizes. Our distant sample is sparsely distributed and shows
     no sign of belonging to a common substructure.}
   \label{fig:pos_xyz}
\end{figure}

\begin{figure*}
  \centering
  \includegraphics[width=14cm, height=12cm]{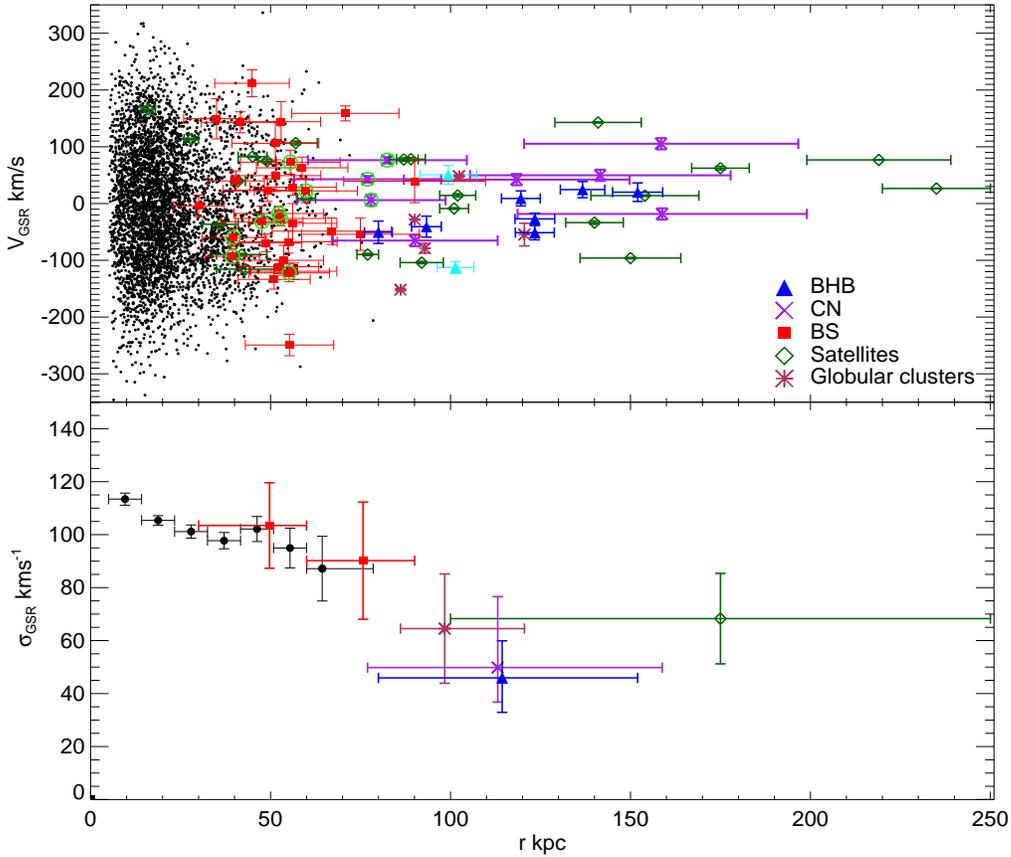}
  \caption[]{\small \textit{Top panel:} Galactocentric velocity as a
    function of radius. The black points are BHB stars selected by
    \cite{xue11} from SDSS DR8. The red squares are BS stars. The blue
    triangles are BHB stars and the cyan triangles are 2 distant field
    BHBs classified by \cite{clewley05}. The purple crosses are CN
    stars. The green circles indicate BS and CN stars that likely
    belong to the Sgr stream. The green diamonds are the Milky Way
    satellite galaxies (excluding Leo I at 250 kpc) and the red
    astericks are the distant Milky Way globular clusters. \textit{Bottom
    panel:} The velocity dispersion profile of field halo stars. The
    colour coding for the different stellar populations is the same as
    the above panel. The migration of stars between the two BS star
    bins due to distance uncertainties is included in their velocity
    dispersion errors. The error in the velocity dispersion for the
    distant BHB stars includes the additional uncertainty of
    excluding/including borderline BHB/BS stars. We also incorporate
    the additional uncertainty of including/excluding possible Sgr
    stream members in the velocity dispersion of the CN stars.}
   \label{fig:sigma}
\end{figure*}

In this section, we look at the distribution of our halo stars in both
position and (line-of-sight) velocity space. We convert our observed
heliocentric velocities to a Galactocentric frame by assuming a
circular speed\footnote{\textbf{We ensured that our results are not strongly
  affected by our choice of the solar reflex motion. Our main results
  are unchanged if we adopt values in the range 180-280 km s$^{-1}$.}} of 240 km s$^{-1}$ (recently revised upwards, see
\citealt{reid09}, \citealt{bovy09}, \citealt{mcmillan11} and \citealt{deason11a}) at the
position of the Sun ($R_0=8.5$ kpc) with a solar peculiar motion
($U,V,W$) = (11.1, 12.24, 7.25) km s$^{-1}$ (\citealt{schonrich10}).

Fig. \ref{fig:sgr} plots the positions of our distant halo stars on
the sky. The top panel shows the right ascension (RA), declination
(DEC) distribution (in J2000) and the middle and bottom panels show
distance and Galactocentric velocity as a function of right
ascension. Our 7 BHB stars are widely distributed over the sky and
show no signs of clustering. The BS stars (and some CN type stars),
however, do show signs of sharing common position and velocity
features. The black lines depict the approximate tracks in position
and velocity space of the Sgr leading and trailing arms. The green
circles indicate stars which coincide with the Sgr streams both in
position, distance and velocity. We identify 8 BS stars that possibly
belong to the Sgr stream. This is unsurprising given the distances of
these stars ($\sim 50$ kpc) and the fact that many of the SDSS
overlaps coincide with this stellar halo stream (see shaded regions in
the top panel of Fig. \ref{fig:sgr}). We also identify one BHB star
that may be associated with Sgr \textit{if} we have classified it
incorrectly. This star (85-TARG33) has an evidence ratio $K=158$ based
on BHB and BS star template fitting and a BHB model is still strongly
favoured if the number ratio prior of BHB-to-BS stars is taken into
account ($\tilde{K}=30$). Thus, we still consider this a genuine
distant BHB star. There are 3 CN stars that likely belong to the Sgr
leading arm. While their distances are slightly too high, given the
uncertainties, we still believe that they could possibly belong to the
Sgr stream. We note that many of the halo carbon stars within $D<100$
kpc have been identified as belonging to the Sgr stream (see
e.g. \citealt{ibata01}), so it is unsurprising that some of our
nearest CN stars in our sample may also belong to this overdensity.

In Fig. \ref{fig:pos_xyz} we show the $x, y$ and $z$ positions of our
distant BHB and CN stars. For comparison, we show the spatial
distribution of the \cite{xue11} sample of BHB stars with distances $D
< 60$ kpc. Our sample of distant stellar halo stars are distributed
over a wide range of distances and sky area.  In summary, whilst some
BS (and CN) stars in our sample likely belong to a known stellar halo
substructure, there is no evidence that our distant BHB and CN stars
belong to a \textit{common} overdensity.

\section{Velocity dispersion profile}

In this section, we analyse the kinematics of our distant stellar halo
sample. The top panel of Fig. \ref{fig:sigma} shows the Galactocentric
velocities of stellar halo stars as a function of Galactocentric
distance. The black points are the BHB spectroscopic sample compiled
by \cite{xue11} from SDSS DR8; these stars probe out to $r \sim 60$
kpc. The red squares are our BS stars. The blue triangles are our 7
distant BHB stars. We also show the 2 distant (field) BHB stars
identified by \cite{clewley05} with the cyan triangles. The purple
crosses are our 8 distant CN type stars. The green circles indicate
the BS and CN stars, which likely belong to the Sgr stream. We also
show the (classical and ultra-faint) Milky Way satellite galaxies and
the distant (beyond 80 kpc) Milky Way globular clusters by the green
diamonds and maroon asterisks respectively. We note that there are now
31 tracers beyond 80 kpc. This is a significant improvement to the 10
tracers (5 globular clusters, 4 satellite galaxies and 1 halo star) used by \cite{battaglia05}.

The bottom panel gives the velocity dispersion profile for stellar
halo stars\footnote{\textbf{We note that when deriving the velocity
    dispersions, the velocity errors have been
    subtracted in quadrature}}. We also give the dispersion for satellite galaxies within
$100 < r/\mathrm{kpc} < 250$ by the green error bar. We find the
striking result that the velocity dispersion is \textit{remarkably low
  at large distances}, dropping to $\sigma \sim 50-60$ km s$^{-1}$
between $100 < r/\mathrm{kpc} < 150$. The outer parts of the stellar
halo seemingly comprise a cold and tenuous veil. A less drastic
decline, but still consistent within the errors, is seen in the
satellite galaxies and globular clusters in this radial regime, as
first noticed by \cite{battaglia05}.

\begin{figure}
 \centering
 \includegraphics[width=8.5cm, height=8.5cm]{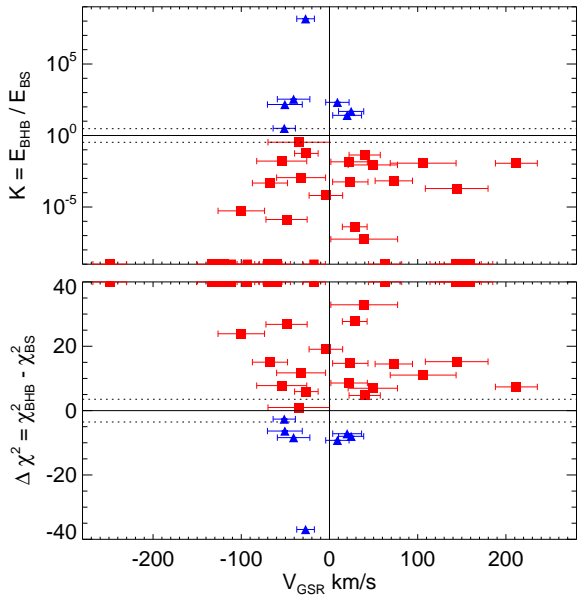}
 \caption{\small \textit{Top panel:} Bayes factor as a function of
   Galactocentric radial velocity. Red squares and blue triangles
   denote BS and BHB stars respectively. The dotted lines at $K=3$ and
   $K=1/3$ indicate a `substantial' strength of evidence favouring a
   BHB or BS model, respectively. BS stars with a very low Bayes factor
   ($< 10^{-9}$) are shown at the $10^{-9}$ level. \textit{Bottom panel:} The
   difference between the minimum Chi-square BHB model and the minimum Chi-square BS
   model as a function of Galactocentric radial velocity. The dotted
   lines indicate 68\% confidence boundaries (assuming 3 degrees of
   freedom). BS stars with $\Delta\chi^2 >40$ are shown at the
   $\Delta\chi^2 =40$ level.}
 \label{fig:pbhb}
\end{figure}

Owing to the small number statistics in the outermost bin for BHB
stars ($\sim 7$ stars), we consider the additional uncertainty in this
measurement if borderline BHB/BS stars are included or excluded. In
Fig. \ref{fig:pbhb}, we show Galactocentric velocity against the
evidence ratio ($K$) of BHB versus BS template model fits (top panel)
and the difference in Chi-square between the best-fitting BHB and BS
models ($\Delta\chi^2$) (bottom panel). We classify stars according to
the evidence ratio with BHB stars have $K > 1$. However, BHBs with $1
< K < 3$ or BSs with $1/3 < K <1$ do not \textit{strongly} favour
particular model templates. The top panel of Fig. \ref{fig:pbhb} shows
that there are two borderline cases (1 BHB, 1 BS). These stars have
correspondingly small differences between the Chi-square values of
their best fitting BHB and BS templates (see bottom panel). We
consider the effect on our results if these borderline cases have been
misclassified. This additional uncertainty (i.e. whether we classify
BHB stars $K > 1/3, 1, 3$) is included in the uncertainty of the
velocity dispersion in Fig. \ref{fig:sigma}. We also note that the
uncertainty on the CN star velocity dispersion takes into account the
effect caused by the inclusion/exclusion of the 3 possible Sgr leading
arm members. Thus, our results are robust to the treatment of
borderline cases.

The observed cold outer stellar halo has implications for the total
mass of our Galaxy, which we now compute using the tracer mass
estimator of \cite{watkins10}. We apply the estimator to distant
stellar halo stars with $r > 50$ kpc. This distant sample comprises
144 BHB and BS stars between $50-90$ kpc and 17 BHB and CN stars
between $80-160$ kpc. The \cite{watkins10} mass estimators assume that the tracer population has a scale-free density and moves in a scale-free potential in the region of interest. The total mass within the outermost tracer
depends on 3 unknown parameters: the tracer density profile slope
($\alpha$), the tracer velocity anisotropy ($\beta=1-\sigma^2_t/\sigma^2_r$)
and the slope of the mass profile ($\gamma$). The mass within the
outermost tracer ($r_{\rm out}$) is computed using equation 16 in \cite{watkins10}\footnote{Note that \cite{watkins10} label $\gamma$ as the tracer density profile, and $\alpha$ as the slope of the mass profile. We use the opposite convention in this work}:
\begin{equation}
\label{eq:mest}
M=\frac{C}{G}\langle v_r^2r^\gamma \rangle, \: \: \:
C=\left( \gamma+\alpha-2\beta \right)r_{\rm out}^{1-\gamma}
\end{equation}

\begin{figure*}
  \centering
  \includegraphics[width=15cm, height=12cm]{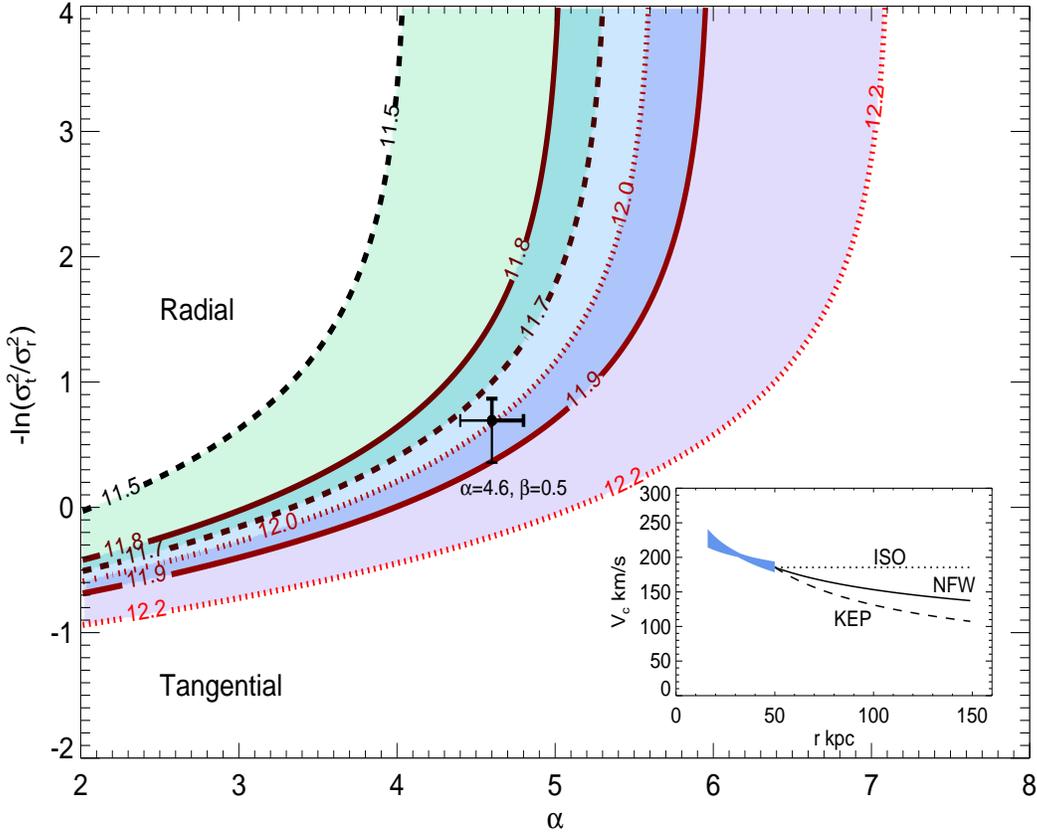}
  \caption[]{\small The contours show the estimated mass within 150
    kpc for different potential profiles. The mass increases from black ($10^{11.5}M_\odot$) to red
    ($10^{12.2}M_\odot$). The solid contours are the masses for NFW haloes
    ($\gamma=0.55$) and the dotted and dashed lines are for isothermal
    ($\gamma=0$) and Keplerian ($\gamma=1$) models respectively. Only
    models consistent with the recently measured mass within 50 kpc
    ($M(50)=4.2 \pm 0.4 \times 10^{11}$; \citealt{deason12}) are
    shown. The allowed regions (for combinations of anisotropy, density and mass) for Keplerian, NFW and isothermal
    models are indicated by the green, blue and purple filled regions. The
    black error bar indicates the density profile and velocity
    anisotropy for tracers within 50 kpc (\citealt{deason11b};
    \citealt{deason12}). If the tracers have moderate density fall-off
    ($\alpha < 5$) and radial orbits ($\sigma^2_t/\sigma_r^2 <1$) then
    the mass within 150 kpc is less than $10^{12}M_\odot$. The inset
    panel shows the the circular velocity curve for
    example isothermal (dotted line), NFW (solid line) and Keplerian
    (dashed line) models. The blue shaded region indicates the
    circular velocity profile measured by \cite{deason12} within 50 kpc.}
   \label{fig:mass}
\end{figure*}

In this radial regime
($50 < r/\mathrm{kpc} < 150$), we can take $\gamma \approx 0.55$,
which is appropriate for Navarro-Frenk-White (NFW; \citealt{nfw}) type
haloes beyond the scale radius, $r>r_{s}$ (see
\citealt{watkins10}). We also consider isothermal ($\gamma=0$) and
Keplerian ($\gamma=1$) halo models. We show
the total mass as a function of the remaining unknown parameters in
Fig. \ref{fig:mass}. We discard any models that are inconsistent (by more
than 2$\sigma$) with the recent mass estimate by \cite{deason12}
within 50 kpc: $M(50)=4.2 \pm 0.4 \times 10^{11}$.  We have checked
that our mass estimates are not significantly affected by the
inclusion/exclusion of borderline cases (i.e. misidentified BHB/BS
stars and Sgr CN stars). The uncertainties due to the unknown tracer
density fall-off and orbital properties dominate over observational errors and small number statistics. 

The mass within 150 kpc ranges from $10^{11.5}-10^{12.2}M_\odot$ depending
on the adopted potential profile, tracer density fall-off and orbital
structure. Massive haloes ($> 10^{12}M_\odot$) require an isothermal profile
between 50-150 kpc and a tracer density profile that falls off very
quickly ($\alpha > 5$), and/or tangentially biased tracer orbits
($\sigma^2_t/\sigma^2_r > 1$). The latter possibility was advocated by
\cite{battaglia05} to explain the apparent decline in velocity
dispersion at large distances whilst \cite{dehnen06} advocated the
former. The allowed range of masses for NFW ($M(150) =6-8 \times
10^{11}M_\odot$) and Keplerian ($M(150) =3.5-5 \times 10^{11}M_\odot$) models are
all less than $10^{12}M_\odot$ over a large range of tracer
properties.

\section{Discussion}
\label{sec:disc}
\subsection{Unrelaxed stars?}

Commonly used mass estimators, such as those derived in
\cite{watkins10}, implicitly assume that the kinematic tracers are
relaxed. How applicable is this assumption for the distant BHB stars
in our sample? Recently, \cite{deason11b} showed that the stellar
halo, as traced by BHB stars, is relatively smooth within $r < 50$
kpc. On the other hand, \cite{bell08} find a much lumpier stellar halo
when traced by main-sequence turn-off stars. While BHB tracers seem to
be the most relaxed of any stellar halo tracer, it is not easy to
extrapolate the properties of the inner stellar halo to the
outskirts. In fact, the much longer dynamical time scales at larger
radii suggest that the outer reaches of the stellar halo are likely
dominated by unrelaxed substructure. Can we then trust our dynamical
mass estimators in this distant radial regime?

\cite{wilkinson99} estimated how correlations in phase-space may
effect mass estimates by considering the (somewhat extreme) case in
which all their data lie along two streams.  The authors found that
mass is systematically underestimated by 20-50 per cent in this
special case.  More recently, \cite{yencho06} considered a more
general case. They applied standard mass estimators to samples of
tracers drawn from random realisations of galaxy haloes containing
levels of substructure consistent with current models of structure
formation. The authors found distortions in their mass estimates at
the level of 20 per cent.

\begin{figure}
 \centering
 \includegraphics[width=8cm, height=6.4cm]{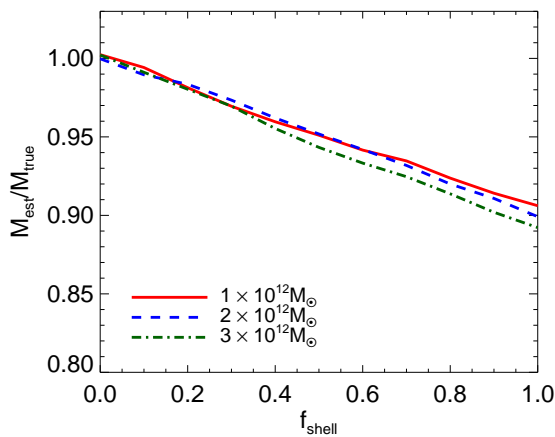}
 \caption{\small The bias in our mass estimates introduced by the
   presence of shells at large radii. We show examples of three
   different mass haloes ($1,2,3 \times 10^{12}M_\odot$) by the solid
   red, dashed blue and dot-dashed green lines, respectively. In each
   case, smoothly distributed tracers with density profile
   $\alpha=3.5$ and radially anisotropic orbits ($\beta=0.5$) are
   embedded in NFW-type haloes ($\gamma=0.55$). When the mass
   estimator is applied we ensure that 10 per cent of the sample has
   $r > 80$ kpc, which is the same fraction as the observed sample
   (there are 17 stars (out of 161) beyond 80 kpc). A fraction of stars
   beyond $80$ kpc ($f_{\rm shell}$) are given a low radial velocity
   distribution (with $\sigma_r \approx 50$ km s$^{-1}$) to mimic the
   presence of shells superimposed on a smooth distribution of
   stars. The presence of shells at large distances causes an
   underestimate in the mass by up to 10 per cent.}
 \label{fig:shells}
\end{figure}

A drop in radial velocity dispersion could signify that our distant
stars belong to shell-like structures. Numerical simulations show that
when a smaller galaxy collides with a larger system, a series of
shell-like structures can form (e.g. \citealt{quinn84}). The liberated
stars from the collision follow very radial orbits with similar
energies. A large enhancement of these stars at apocentre can thus
build up into shell structures. The velocity structure in these shells
is coincident with the systemic velocity of the host galaxy, with a
very low dispersion. We quantify the effect of shells on our mass
estimates by generating fake distributions of stars drawn from
appropriate power-law distribution functions
(cf. \citealt{watkins10}). We consider smoothly distributed tracers
($\alpha=3.5$, between $50 < r/\mathrm{kpc} < 150$) with radially
anisotropic orbits ($\beta=0.5$) embedded in NFW type haloes
($\gamma=0.55$). We then superimpose stars at large distances ($r >
80$ kpc) with low radial velocities (with $\sigma_r \sim 50$ km
s$^{-1}$) onto this smooth distribution. The mass estimators used in
the previous section (see eqn. \ref{eq:mest}) are then applied to this test data set, for which
we assume $\alpha$, $\gamma$ and $\beta$ are known. In our
observational sample, 10 per cent of the stars are in the outer mass
bin ($r > 80$ kpc). We ensure the same fraction of stars are in this
outer mass bin when we estimate the mass of the fake data set. In Fig. \ref{fig:shells}, we show the results of this exercise, in which
we vary the fraction of stars beyond $r=80$ kpc that belong to shells
from 0-100 per cent. We show three examples of halos with masses within
150 kpc of $1 \times 10^{12}M_\odot$ (solid red), $2 \times
10^{12}M_\odot$ (dashed blue) and $3 \times 10^{12}M_\odot$
(dot-dashed green) respectively. This simple calculation shows that
the mass is underestimated by up to 10 per cent if \textit{most} of
the stars beyond $80$ kpc belong to shells. A larger sample of stars
beyond $80$ kpc is needed to to test whether or not this drop in
radial velocity dispersion is indeed caused by the presence of shells.

In summary, it is unlikely that substructure is strongly biasing our
mass estimates within $r=150$ kpc. Tidal streams are excluded by the
the lack of phase-space correlation in our distant BHB and CN star
sample. It seems reasonable to assume the presence of substructure at
large radii introduces an additional uncertainty in our mass estimate
at the level of 20 per cent. In particular, our mass estimates may be
biased low (by up to 10 per cent) if many of our distant stars belong to
shells. Even so, this is a negligible contribution compared to the
systematic uncertainty in the tracer properties such as density and
anisotropy.

\subsection{Tangential anisotropy and/or rapid tracer density fall-off?}

The cold radial velocity dispersion at large radii may be caused by
the dominance of tangential motions. But is there physical
justification for this picture? Current theories predict that the
stripped stellar material from accreted satellite galaxies make up the
bulk of the stellar halo (e.g. \citealt{bullock05}; \citealt{cooper10}). Recently, other formation scenarios have
been suggested for the inner stellar halo (i.e. \textit{in situ} formation,
see e.g. \citealt{zolotov09}; \citealt{font11}), but accretion seems
the only likely scenario in the outer regions. Model stellar haloes
built up from the accretion of satellites predict \textit{strongly
radial orbits} (e.g. \citealt{diemand07};
\citealt{sales07}). Observational constraints from the solar
neighbourhood and inner ($< 50$ kpc) stellar halo find radially biased
orbits ($\sigma^2_t/\sigma^2_r \sim 0.5$, e.g \citealt{smith09};
\citealt{deason12}) and thus seem to concur with theoretical
predictions. The fact that the radial anisotropy is only expected to
increase with radius further contradicts the notion of tangentially
biased orbits in the radial regime $100 <r/\mathrm{kpc} < 150$. Thus,
given current observational constraints and theoretical predictions,
it is unlikely that tangential anisotropy can explain the presence of
such a cold radial velocity dispersion.

Another possibility is that the density profile of the stellar halo
falls sharply at large distances. While the stellar halo density
profile within 50 kpc has been studied extensively
(e.g. \citealt{bell08}; \citealt{juric08}; \citealt{deason11b};
\citealt{sesar11}), constraints at larger radii are rare. In the
radial range of our distant stars ($100 < r/\mathrm{kpc} < 150$), we
are probing the tail of the stellar distribution and it is reasonable
to assume that the stellar density may be falling off very rapidly. In
fact, \cite{dehnen06} showed that a sharp decline in velocity
dispersion can be produced by radially anisotropic
($\sigma^2_t/\sigma^2_r \sim 0.5$) tracers with a truncated density
distribution ($r_t =160$ kpc) embedded within a $1.5 \times
10^{12}M_\odot$ halo. Does such a cold velocity dispersion then
signify that we are probing the edge of the stellar halo?
Spectroscopic programmes targeting halo stars beyond 150 kpc will be
vital in order to address this point.

It is important to bear in mind that a low (albeit less extreme)
velocity dispersion is also seen in the satellite galaxy
population. There is no requirement that the stellar halo stars and
satellite galaxies should share a similar density profile and/or
velocity anisotropy. In fact, the distribution of satellite galaxies
in our own Galaxy and M31 is believed to be much shallower than
stellar halo stars (e.g. \citealt{watkins10}). Furthermore, the proper
motions of the classical Milky Way satellites suggest that they have
tangentially biased orbits, whereas numerical simulations predict that
stellar halo stars have strongly radial orbits. The fact that
\textit{both} of these halo populations have a cold radial velocity
dispersion beyond 100 kpc suggests that there is a common cause,
namely part (if not all) of the decline must be related to the mass
profile.

\subsection{A low mass, high concentration halo?}

It is clear from Fig. \ref{fig:mass} that, discounting a stellar
population with tangential velocity anisotropy ($\sigma^2_t/\sigma^2_r
> 1$) and/or a rapid decline in density between $50-150$ kpc ($\alpha
> 5$), the mass within 150 kpc is less than $10^{12} M_\odot$ and
probably lies in the range $M(<150 \mathrm{kpc}) = (5-10) \times
10^{11} M_\odot$. In particular, if we extrapolate measurements of
tracer density and anisotropy within $50$ kpc to $150$ kpc, namely
$\alpha \sim 4.6$ and $\beta=0.5$, we find that $M(<150 \mathrm{kpc})
\sim 7 \times 10^{11} M_\odot$ (assuming an NFW halo with $\gamma=0.55$). Given that the total mass within 50
kpc is $\sim 4 \times 10^{11}M_\odot$, our results suggest that
\textit{there may be little mass between 50 and 150 kpc}. Is most of
the dark matter in our Galaxy therefore highly concentrated in the centre?
\cite{deason12} recently found that the dark matter profile of our
Galaxy within 50 kpc is highly concentrated ($c_{\rm vir} \sim
20$). Our results presented here for the outer halo are in agreement
with the deductions from the inner halo.

Our findings also agree with several other recent studies suggesting
that the Milky Way may be less massive than previously thought
(e.g. \citealt{battaglia05}; \citealt{smith07}; \citealt{xue08}).
This has important repercussions for numerical simulations attempting
to reproduce Milky Way type galaxies, where often the biggest scatter
in halo properties comes from their total masses. For example,
\cite{wang12} recently showed that the number of massive Milky Way
subhaloes strongly depends on the virial mass of the halo (see also \citealt{vera12}). The lack of
massive satellites in our Galaxy (the `missing massive satellites'
problem, \citealt{kolchin12}) could simply indicate a less massive
halo ($M_{200} < 10^{12}M_{\odot}$). However, a low halo mass may
prove problematic for semi-analytic models which attempt to match the
observed Tully-Fisher relation and galaxy stellar mass function
(e.g. \citealt{cole00}; \citealt{croton06}). Generally, agreement with
observations requires that the virial velocity ($V_{200} \sim 150$ km
s$^{-1}$ for $M_{200} \sim 10^{12}M_\odot$) and rotation speed of the galaxy
are comparable (i.e. $V_{200} \sim V_{\rm rot}$). Thus, a rotation
speed of 220-250 km s$^{-1}$ for the Milky Way disc and a low halo
mass is at odds with these semi-analytic models.

A low mass Milky Way halo also has implications regarding the origin
of several satellite galaxies. Leo I would almost certainly be unbound
with a distance of 250 kpc and a substantial radial velocity of
$V_{\rm GSR} \sim 180$ km s$^{-1}$. In addition, recent proper motion
measurements of the Small and Large Magellanic Clouds (SMC/LMC)
(\citealt{kallivayalil06a}; \citealt{kallivayalil06b}) suggest that
the total velocities of these clouds approach the escape speed of the
Milky Way. \cite{besla07} showed that under the assumption of a Milky
Way with virial mass $M_{\rm vir} \sim 10^{12}M_\odot$, the LMC and SMC are
on their first passage about the Milky Way. The current results seem
to increase the likelihood that the SMC/LMC are unbound.

\section{Conclusions}

We have built a sample of distant stellar halo stars beyond $D > 80$
kpc with measured radial velocities. Our sample consists of relatively
old BHB stars and intermediate age AGB stars. We target BHB stars in
the magnitude range $20 <g <22$ using multi-epoch SDSS
photometry. Using follow-up spectroscopic observations with the
VLT-FORS2 instrument we obtained radial velocities for 38 A-type
stars. We distinguish between BHB and BS stars by fitting noise-free
BHB and BS templates built from high S/N SDSS spectra. Our sample
comprises 7 BHB stars in the radial range $80 < r_{\rm
BHB}/\mathrm{kpc} < 150$ and 31 less distant BS stars with $30 <
r_{\rm BS}/\mathrm{kpc} < 90$.  In addition, a sample of 4 distant
cool carbon stars with measured radial velocities was compiled from
the literature. We measured radial velocities for an additional 4
carbon stars using the WHT-ISIS instrument. Our 8 carbon stars span a
distance range $80 < r_{\rm CN} < 160$ kpc and are a useful,
complementary sample to the older BHB population. We summarise our
conclusions as follows:

\medskip
\noindent
(1) The velocity dispersion of the distant BHB and CN stars is
surprisingly low with $\sigma_{\rm GSR} \sim 50-60$ km s$^{-1}$. The
outer parts of the stellar halo may be likened to a cold veil.

\medskip
\noindent
(2) Although, a significant number of the BS and CN stars (8 and 3
respectively) are coincident with the Sagittarius (Sgr) stream, our
sample of BHB stars (and the remaining CN stars) are widely
distributed across the sky and show no evidence that they belong to a
common substructure. However, such a low radial velocity dispersion
could be caused by the presence of shells at large radii.

\medskip
\noindent
(3) The observed low velocity dispersion profile is robust to the
inclusion/exclusion of borderline BHB candidates and CN star Sgr
members. The velocity dispersion of satellite galaxies in a similar
radial range ($ 100 < r/\mathrm{kpc} < 250$) is also low, $\sigma_{\rm
  GSR} \sim 70$ km s$^{-1}$. \cite{battaglia05} first noted this
decline in radial velocity dispersion for distant satellite galaxy and
globular cluster tracers. We now confirm that a similar, if not more
extreme, cold velocity dispersion is seen in distant stellar halo
stars.

\medskip
\noindent
(4) The implications for the total mass of our Galaxy depends on the
(unknown) density profile and velocity anisotropy of the tracer
population. However, discounting a stellar population with a
tangential velocity bias ($\sigma^2_t/\sigma^2_r > 1$) and/or a rapid
tracer density fall-off ($\alpha > 5$) between $50-150$ kpc, we find
that the total mass within 150 kpc is less than $10^{12} M_\odot$ and
probably lies in the range $(5-10) \times 10^{11}M_\odot$. Our results
thus suggest that the total mass of our Galaxy may be lower than
previously thought.

\bigskip
\noindent
In this study, we have expanded the sample of tracers out to $ r \sim
150$ kpc --- near the edge of the stellar halo. Larger samples of
tracers are needed to add further statistical weight to our results. We intend
to address this issue over the next few years by combining deep and
wide photometric surveys with 4-10m class telescopes equipped with
moderate/high resolution spectrographs. Ultimately, even with a much
larger number of halo tracers, the uncertainty surrounding the tracer
velocity anisotropy and density needs to be addressed. Over the next
few years it will become vital to \textit{measure} these tracer
properties rather than to infer them.

\section*{Acknowledgements}
AJD thanks the Science and Technology Facilities Council (STFC) for
the award of a studentship. VB acknowledges financial support from the
Royal Society. S.K. acknowledges financial support from the STFC. RC is supported by a Research
Fellowship at Peterhouse College, Cambridge. JP acknowledges support
from the Ram\'on y Cajal Program as well as by the Spanish grant
AYA2010-17631 awarded by the Ministerio of Econom\'ia y
Competitividad. CFPL is currently supported by the Marie Curie Initial
Training Network CosmoComp (PITN-GA-2009-238356. MF acknowledges funding
through FONDECYT project No. 1095092 and BASAL. MGW is currently
supported by NASA through Hubble Fellowship grant HST-HF-51283.01-A,
awarded by the Space Telescope Science Institute, which is operated by
the Association of Universities for Research in Astronomy, Inc., for
NASA, under contract NAS5-26555. EO was partially supported by NSF
grant AST0807498. It is a pleasure to thank Mike Irwin and Paul Hewett
for useful comments and advice. We also thank an anonymous referee for
useful comments.

\label{lastpage}

\bibliography{mybib}

\end{document}